\documentclass[a4paper,11pt]{article}
\usepackage{pos}

\title{Open charm and dileptons from relativistic heavy-ion collisions}

\author*[a,b]{Elena Bratkovskaya}
\author[a]{Taesoo Song}
\author[c]{Pierre Moreau}
\author[b]{Carsten Greiner}

\affiliation[a]{GSI Helmholtzzentrum f\"{u}r Schwerionenforschung GmbH, 
Planckstrasse 1, 64291 Darmstadt, Germany }

\affiliation[b]{Institute for Theoretical Physics, Johann Wolfgang Goethe Universit\"{a}t, Frankfurt am Main, Germany}

\affiliation[b]{Department of Physics, Duke University, 
Durham, North Carolina 27708, USA}

\emailAdd{E.Bratkovskaya@gsi.de}

\abstract{
We study the dynamics of open charm production and the 
dilepton radiation of the semi-leptonic decays of correlated $D\bar D$ pairs
versus the quark-gluon plasma (QGP) radiation and hadronic sources 
in relativistic heavy-ion collisions.
Our study is based on the Parton-Hadron-String Dynamics (PHSD) transport approach
employing a non-perturbative QCD description of the strongly interacting 
quark-gluon plasma (sQGP) in  terms of dynamical quasiparticles and the
EoS based on lattice QCD.
We compare the PHSD results for charm observables with the calculations from  
BAMPS (Boltzmann Approach to Multi-Parton Scatterings) 
which is based on perturbative QCD with massless partons and
interaction cross sections calculated in leading order of the QCD coupling.
We compare the  $p_T$ dependence of the ratio $R_{AA}$ of  $D$-mesons in $A+A$ over
$p+p$ collisions scaled by the  number of binary 
collisions $N_{bin}$ as well as the elliptic flow $v_2$ of $D$-mesons 
calculated within the PHSD and BAMPS at LHC energies.
In other study, based on the PHSD calculations we find that the  dileptons from correlated 
$D-$meson semi-leptonic decays dominate the 'thermal' radiation from
the QGP in central Pb+Pb collisions at the intermediate masses 
($1.2 < M < 3$ GeV) for higher  invariant energies 
However, for invariant energies $\sqrt{s_{\rm NN}} < 40$ GeV 
the QGP radiation overshines the contribution from $D,{\bar D}$
decays such that one should observe a rather clear signal 
from the partonic dilepton radiation.
This finding provides promising perspectives to measure the QGP radiation
in  the dilepton experiments at RHIC BES and the future FAIR/NICA facilities.
}

\FullConference{%
  HardProbes2020\\
  1-6 June 2020\\
  Austin, Texas}

\begin{document}
\maketitle

Heavy-ion collisions provide a unique possibility to study the properties 
of strongly interacting matter under extreme conditions. 
The heavy quarks ($D$-mesons) and electromagnetic radiation 
(real photons and dileptons) are considered to be the most sensitive
observables which provide access to 
the interior of HIC collisions from early to the final stages of the reaction.
The charm  quark ($c {\bar c}$) pairs are produced through initial hard nucleon-nucleon scattering in relativistic heavy-ion collisions. Since the production of heavy flavor requires a large energy-momentum, it can be described 
within perturbative QCD  \cite{Cacciari:2012ny}.
The produced heavy flavor then interacts with the hot dense matter
(of partonic or hadronic nature) by exchanging energy and momentum.
As a result, the ratio of the measured number of heavy flavors in
heavy-ion collisions to the expected number in the absence of
nuclear or partonic matter, which is the definition of $R_{\rm AA}$,
is suppressed at high transverse momentum, and the elliptic flow of
heavy flavor is generated by the interactions in noncentral
heavy-ion collisions~\cite{ALICE:2012ab}.

The theoretical description of the charm energy loss is a challenging 
task since it requires the knowledge about the interactions of charm with 
the medium as well as the proper description of the medium time evolution
and interactions of the internal degrees-of-freedom. In this respect 
microscopic transport approaches are  powerful tools to model
the charm dynamics.
Our study of charm energy loss is based on two comprehensive
transport approaches:  PHSD (Parton-Hadron-String Dynamics)
\cite{Cassing:2009vt,Linnyk:2015rco}
and BAMPS (Boltzmann Approach to Multi-Parton Scatterings) \cite{Xu:2004mz,Uphoff:2014cba}.  
The PHSD is a microscopic off-shell transport approach for the description of strongly interacting partonic and hadronic matter in and out-of equilibrium which is based on the solution of Kadanoff--Baym equations in first-order gradient expansion
\cite{Cassing:2008nn}. The realization of the QGP phase is based on the dynamical
quasiparticle model (DQPM)~\cite{Cassing:2008nn,Berrehrah:2016vzw}
which describes the strongly interacting, i.e. non-perturbative, partonic
medium in terms of off-shell massive quasiparticles (quarks and gluons) 
with broad spectral functions; their properties, i.e. complex self-energies, 
are defined from the QGP thermodynamics in line with the lQCD EoS at finite
temperature $T$ and baryon chemical potential $\mu_B$.
The perturbative QCD cascade BAMPS is based on pQCD degrees-of-freedom, i.e. 
massless quarks and gluons which follow the leading-order pQCD interactions,
including radiative processes in an improved Gunion-Bertsch
approach \cite{Fochler:2013epa} in line with a Boltzmann
collision integral.
The employment of the PHSD and BAMPS transport approaches allows
to explore the charm interactions with a perturbative and non-perturbative 
QCD medium.

\section{Charm production}

The realization of charm dynamics in the PHSD is described in detail in Refs. 
\cite{Song:2015sfa}.  
In the QGP phase, the elastic and quasi-elastic interactions of charm quarks 
with off-shell quarks and gluons are calculated within the effective propagators 
and $T$-dependent couplings from the DQPM 
which allow to explain lQCD data on the temperature dependence of the drag coefficient $D_S$ 
at $\mu_B$=0 \cite{Berrehrah:2016vzw}. 
The PHSD includes the dynamical hadronization of charm quarks through coalescence 
and/or fragmentation (depending on transverse momentum).
In the hadronic phase,  $D$-mesons can interact with baryons and mesons with cross sections calculated in an effective hadronic model based on a lagrangian approach
 with heavy-quark spin symmetry  (cf. Refs. in \cite{Song:2015sfa}).

In BAMPS, the interactions of charm quarks with the pQCD medium include elastic and 
radiative heavy flavor interactions calculated within the improved Gunion-Bertsch approximation \cite{Gunion:1981qs}.
The energy loss of the charm quarks has been extensively studed in 
Refs. \cite{Uphoff:2012gb}.

\begin{figure}[t]
\centerline{
\includegraphics[width=6.5cm]{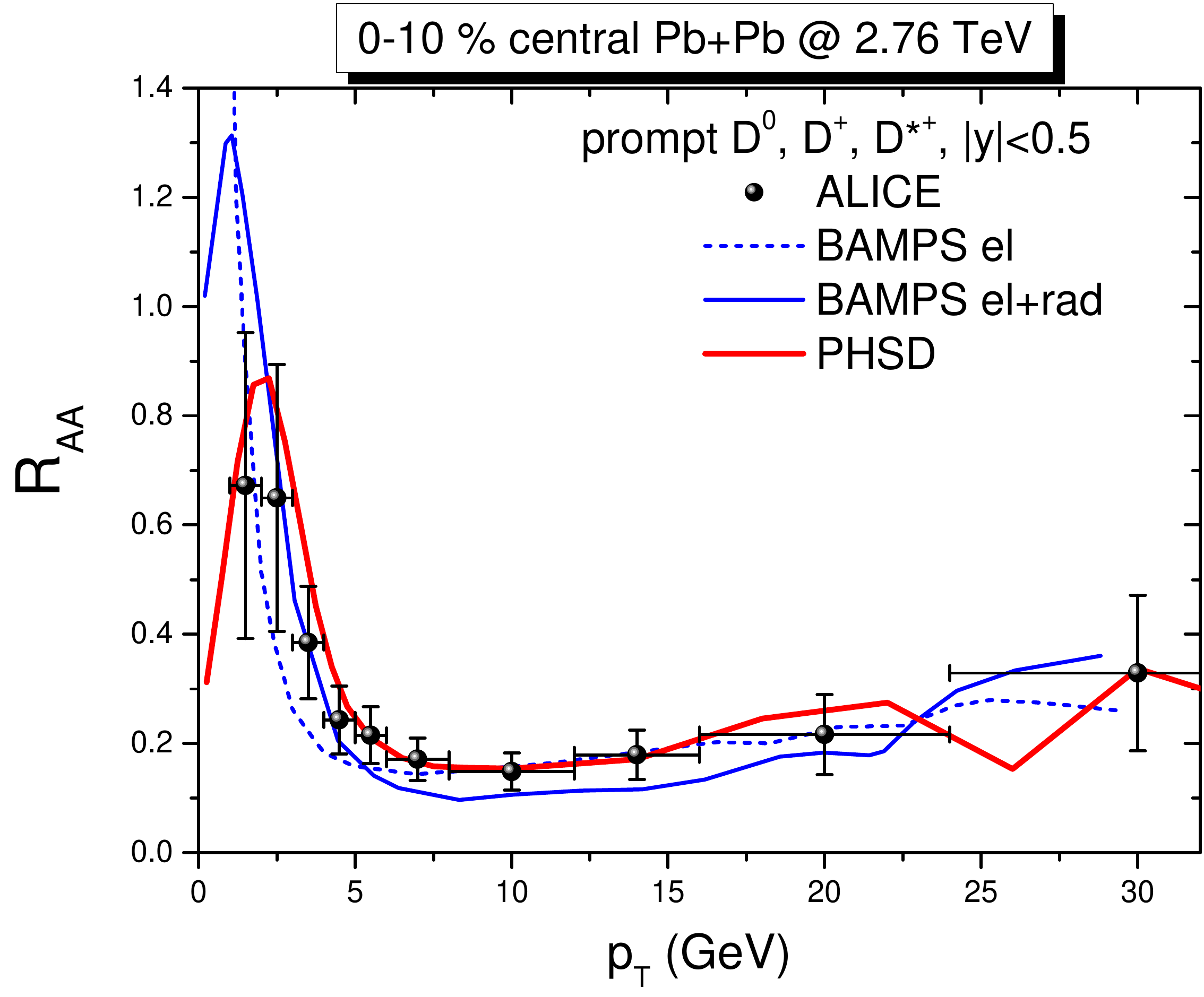}
\hspace*{5mm}\includegraphics[width=6.5cm]{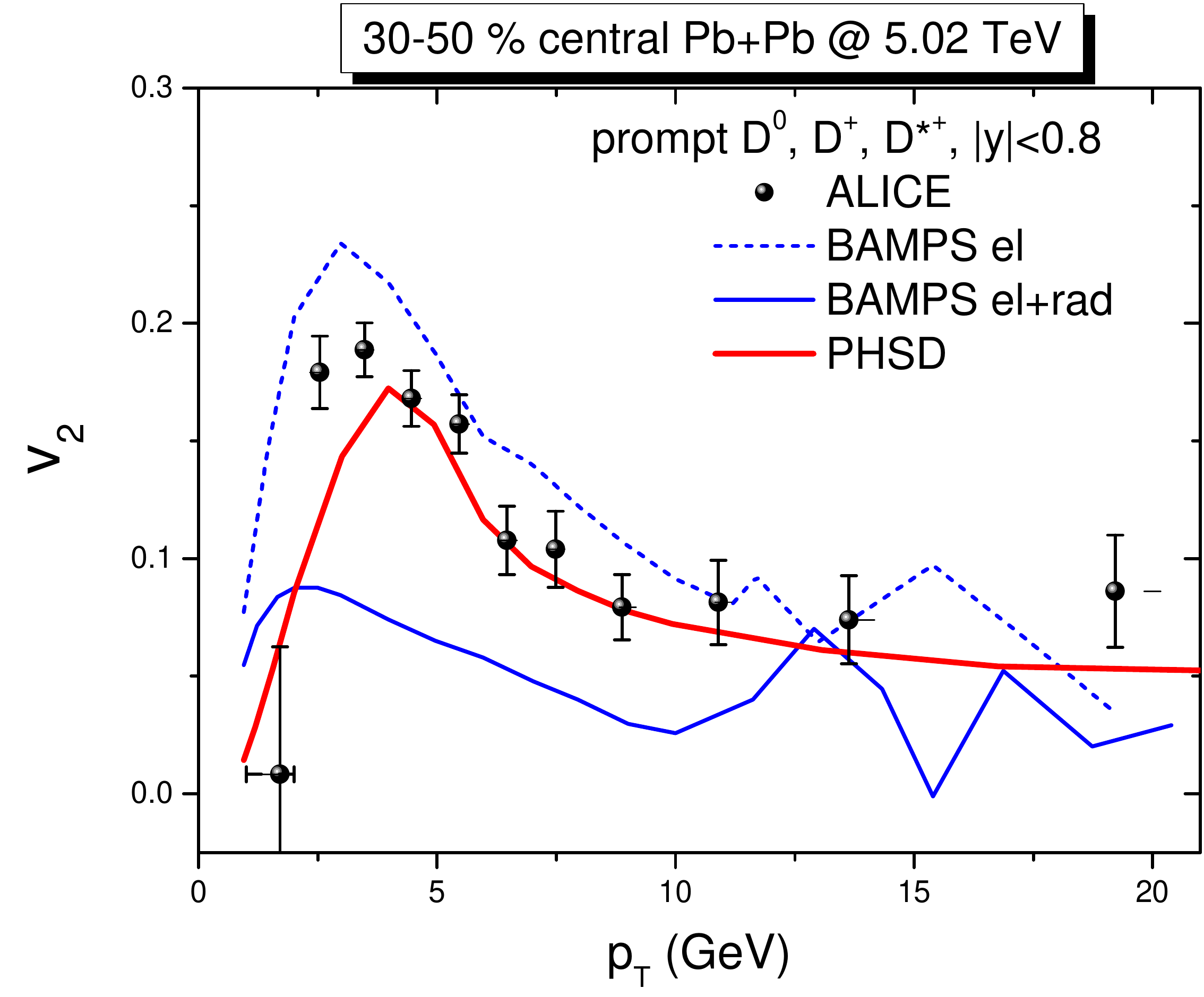}}
\caption{ Left: The ratio  $R_{\rm AA}$ of
$D^0,~D^+$, and $D^{*+}$ mesons within $|y|<0.5$ as a function 
of transverse momentum ${p_{\rm T}}$ in 0-10 \% central Pb+Pb 
collisions at $\sqrt{s_{\rm NN}}=$\ 2.76 TeV compared to 
the experimental data from the ALICE
collaboration \protect\cite{Adam:2015sza}. 
Right: The PHSD and BAMPS results for the elliptic flow $v_2$ 
of $D^0$ mesons within $|y|<0.8$ in 30-50 \%  central Pb+Pb collisions 
at $\sqrt{s_{\rm NN}}=$\ 5.02 TeV compared to the experimental data from the ALICE
collaboration \protect\cite{Acharya:2020pnh}. 
The solid red lines show the PHSD results,
while the blue dashed lines display the BAMPS results including elastic 
scattering only and the blue solid lines -- the BAMPS results with 
elastic and radiative processes.} 
\label{RAA_v2}
\end{figure}

In the left plot of Fig. \ref{RAA_v2} the comparison of  PHSD and BAMPS results 
with ALICE data at 2.76 TeV \cite{Adam:2015sza} for 
the $p_T$ dependence of the ratio $R_{AA}$ 
of $D$-mesons in $A+A$ over $p+p$ collisions, scaled by the  number of binary 
collisions $N_{bin}$, is shown while the right plot of Fig. \ref{RAA_v2} 
shows the elliptic flow $v_2$ of $D$-mesons for both approaches versus 
the ALICE measurement at 5.02 TeV \cite{Acharya:2020pnh}.
The BAMPS results correspond to a scenario with running coupling 
$\alpha_S(Q^2)$ and scaling factor $K=3.5$ for the elastic interactions which is needed for 
the description of the $R_{AA}$ since the pQCD elastic scattering is too small 
to explain the experimental data on $R_{AA}$ for $p_T < 30$~GeV and requires a large 
scaling factor of 3.5. 
Including radiative collisions, on the other hand, the pQCD description in compatible 
with the data, pointing out the potential importance of the radiative processes: 
as seen from the blue solid lines in Fig. \ref{RAA_v2}
the pQCD radiative processes provide extra energy loss, however, their angular dependence
- with a strong forward peak - leads to an underestimation of elliptic flow $v_2(p_T)$. 
Contrary, the non-perturbative QGP interactions with heavy partons and more 
isotropic angular distributions - as in the PHSD - allow to describe 
the $R_{AA}(p_T)$ and $v_2(p_T)$ simultaneously. We note that the gluon 
radiative processes in the DQPM (and correspondingly, in the PHSD) are strongly 
suppressed due to the large mass of thermal gluons contrary to the pQCD based models.

This example shows how sensitive the charm probes are to the properties of the underlying medium.
We mention, that this issue has been studied extensively by the combined efforts of different worldwide groups working on charm dynamics 
\cite{Cao:2018ews,Rapp:2018qla,Xu:2018gux}.

\section{Excitation function of dielectron production }

The electromagnetic probes (dileptons and real photons) are considered 
to provide a powerful probe of the quark-gluon plasma as created in ultra-relativistic 
nuclear collisions. They interact only electromagnetically and 
thus escape to the detector almost undistorted through the dense 
and strongly-interacting medium \cite{Linnyk:2015rco}.
Although the dileptons from  partonic scatterings in the QGP phase
are of  primary interest, there are variety of other dilepton 
channels, i.e. from hadronic decays and interactions in the hadronic  phase
as well as dileptons from semi-leptonic decays of correlated $D\bar{D}$ pairs. 

\begin{figure}[tbh]
\centerline{
\includegraphics[width=6cm]{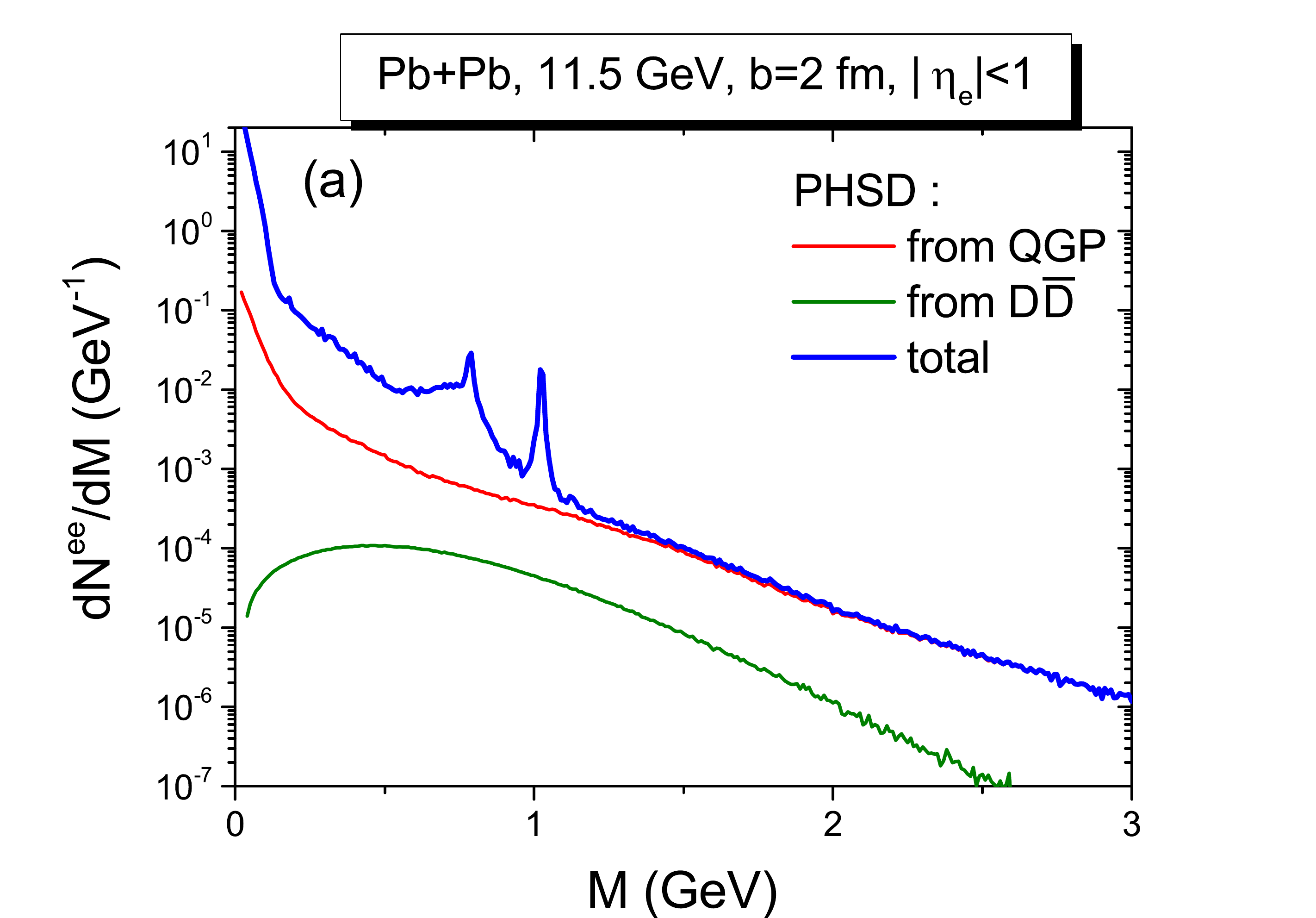}
\hspace*{-10mm}
\includegraphics[width=6cm]{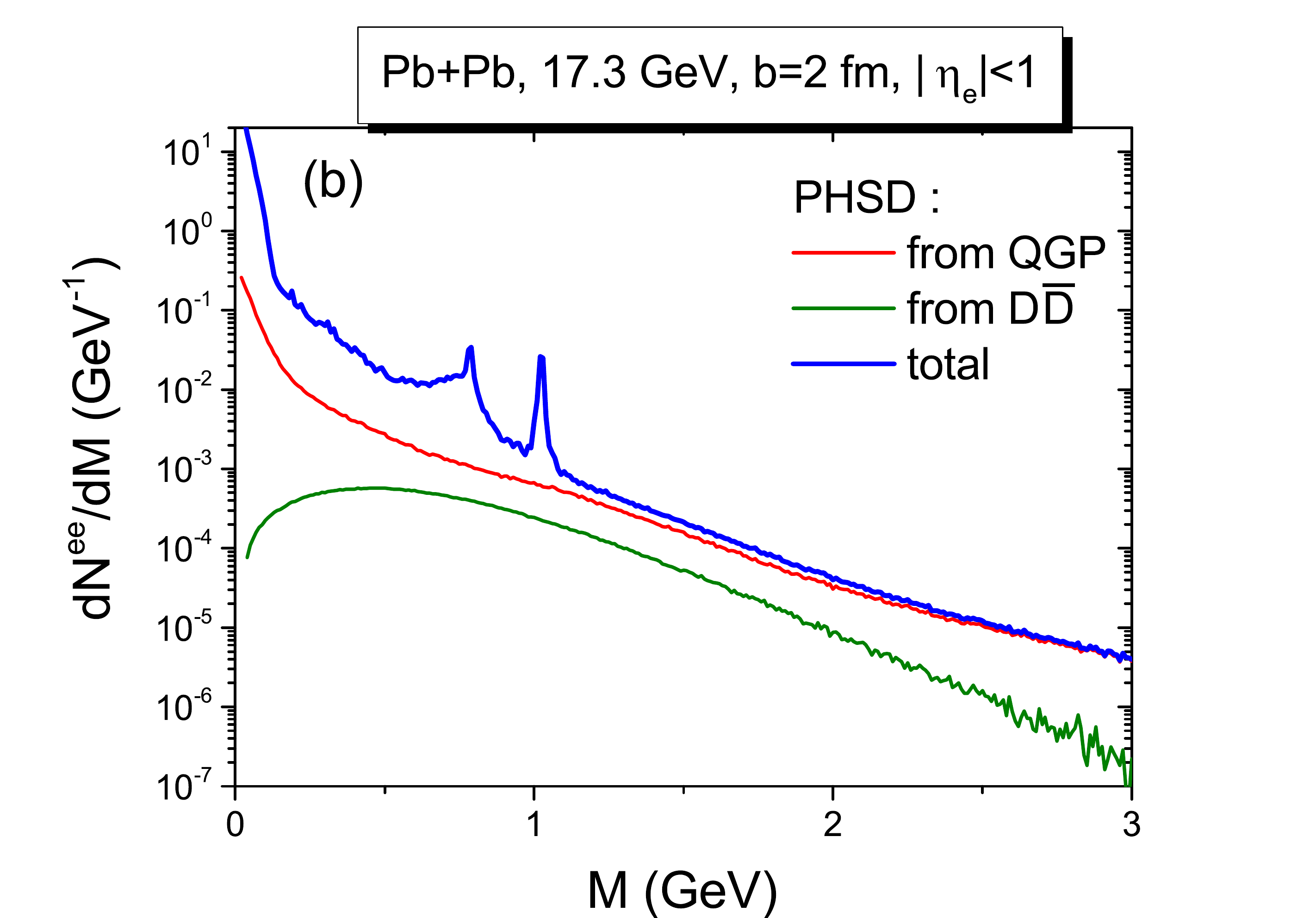}
\hspace*{-10mm}
\includegraphics[width=6cm]{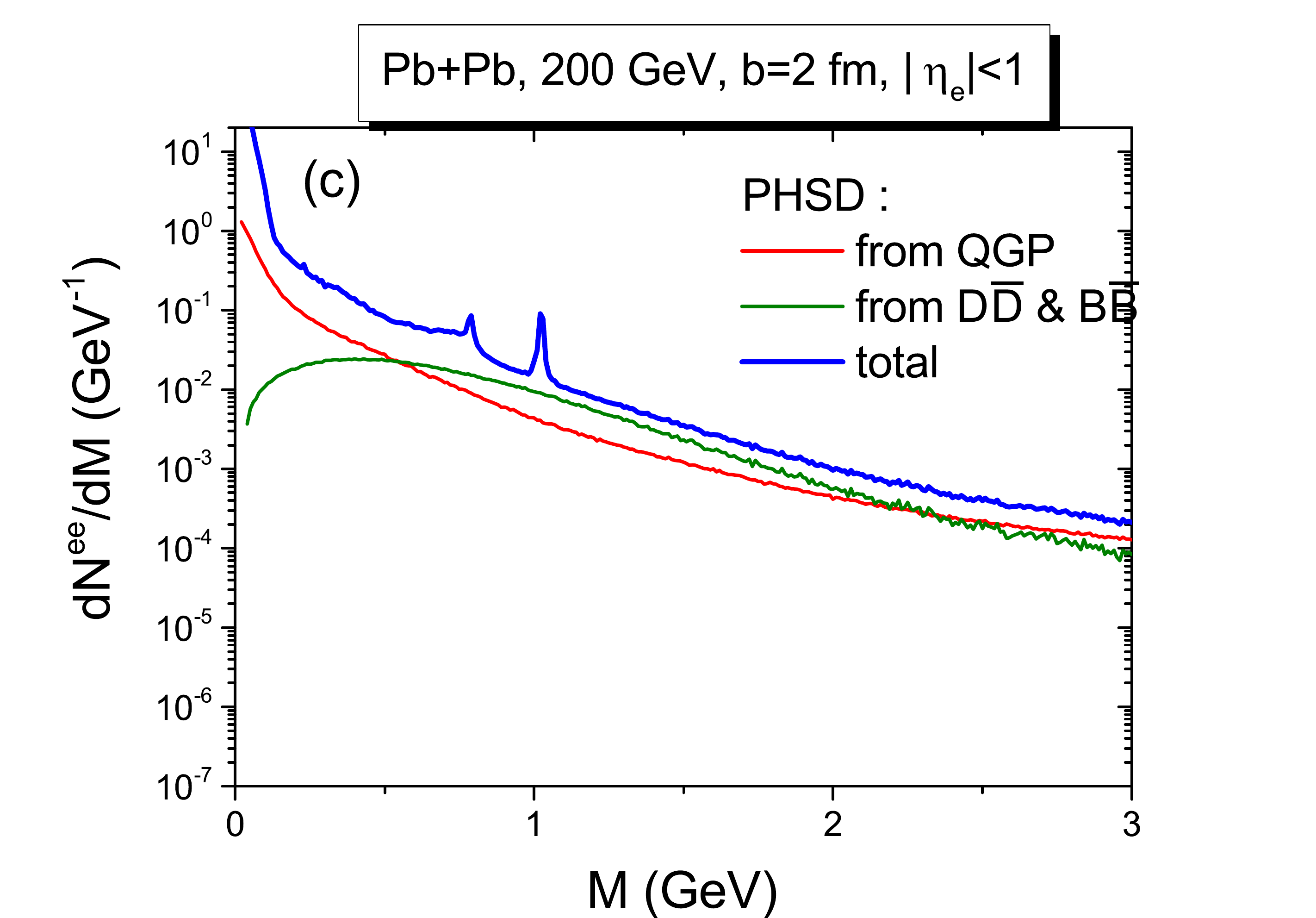}}
\caption{The invariant mass spectra of dileptons from partonic
interactions  (red lines) and from $D\bar{D}$ pairs (green lines)
together with the total dielectron spectrum (blue lines) in central
Pb+Pb collisions at $\sqrt{s_{NN}}$ =  11.5, 17.3, and 200 GeV from
the PHSD at mid-pseudorapidity for the leptons.} \label{fig6}
\end{figure}

In  Ref. \cite{Song:2018xca} we have compared the separate contributions 
from different dilepton channels in central Pb+Pb collisions at 
various energies from $\sqrt{s_{\rm NN}}=$8 to 200 GeV within the PHSD.
In  Fig.~\ref{fig6} we show the contributions from the QGP (red lines),
which is the sum of three partonic channels, i.e.
$q+\bar{q}\rightarrow e^++e^-$, $q+\bar{q}\rightarrow g+e^++e^-$,
and $q(\bar{q})+g\rightarrow q(\bar{q})+e^++e^-$,
and from $D\bar{D}$ pairs (green lines)  with the total dielectron
spectrum (blue lines) at different collision energies for central
Pb+Pb collisions. 
We find that the contribution from the hadronic channels increases only moderately 
with collision energy (in line with the hadron abundances), while the contribution
from the QGP raises more steeply (in line with the enhanced space-time volume of the QGP phase). The contribution from correlated $D\bar{D}$ pairs
is small at low-energy collisions, but becomes more and more important 
with increasing collision energy in competition with the production from the QGP channels.

\begin{figure} [tbh]
\begin{minipage}[l]{6cm}
\includegraphics[width=8 cm]{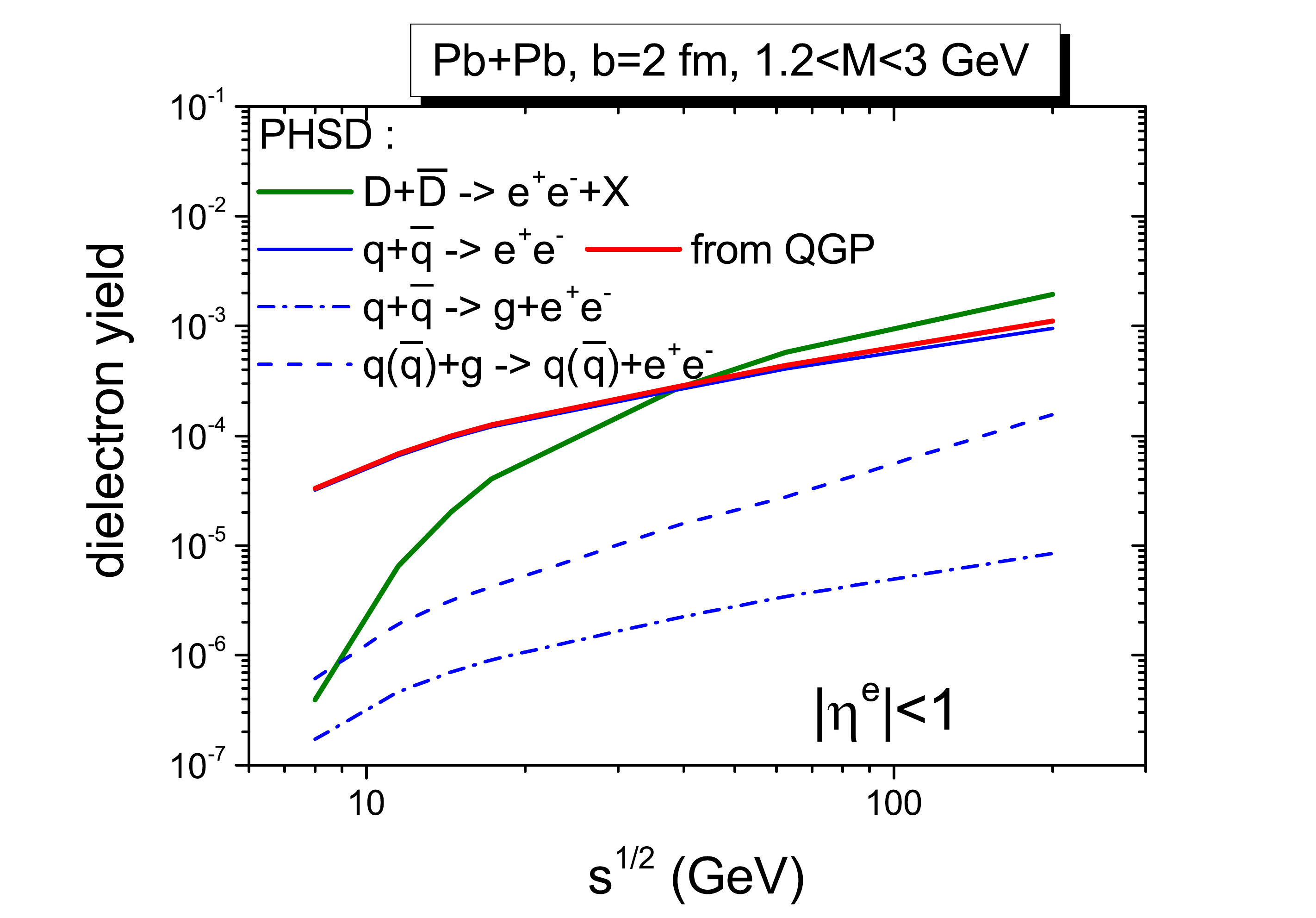}
\end{minipage}\hfill
\begin{minipage}[r]{7.5cm}
\caption{The contributions to intermediate-mass dielectrons (1.2 GeV
$< M <$ 3 GeV)  from $D\bar{D}$ pairs (green lines), different
channels of partonic interactions, $q+\bar{q}\rightarrow e^++e^-$,
$q+\bar{q}\rightarrow g+e^++e^-$, $q(\bar{q})+g\rightarrow
q(\bar{q})+e^++e^-$ (see legend) as a function of $\sqrt{s_{NN}}$
for Pb+Pb collisions at $b=2$ fm for midrapidity leptons. The red
solid line displays the sum of the partonic contributions.}
\label{fig7}
\end{minipage}
\end{figure}

Fig.~\ref{fig7} compares the contribution from $D\bar{D}$ pairs
(green lines) to the QGP contribution for intermediate
mass dileptons (1.2 GeV $< M <$ 3 GeV)  as a function of collision
energy $\sqrt{s_{NN}}$ for Pb+Pb collisions at $b=2$ fm. The figure
clearly shows that the contribution from partonic interactions,
especially from $q+\bar{q}\rightarrow e^++e^-$, dominates the
intermediate-mass range in low-energy collisions. However, the
contribution from $D\bar{D}$ pairs rapidly increases with increasing
collision energy, because the scattering cross section for charm
production grows fast above the threshold energy. 
It overshines the contribution from partonic interactions 
around $\sqrt{s_{NN}} \approx$ 40 GeV and dominates at higher energies.

Thus, our results in Figs.~\ref{fig6} and \ref{fig7} clearly
demonstrate that the window to study partonic matter by dielectrons
at intermediate masses without substantial background from  heavy
flavor decays opens for collision energies $\sqrt{s_{NN}} <$ 40 GeV.
This finding provides promising perspectives for the dilepton experiments 
at RHIC BES and the future FAIR/NICA facilities.

\vspace{5mm}
We acknowledge support by the DFG through the grant CRC-TR 211 
'Strong-interaction matter under extreme conditions' - Project number 
315477589 - TRR 211. 


\end{document}